\documentclass{sigchi}

\usepackage{balance}  
\usepackage{graphics} 
\usepackage{times}    
\usepackage{url}      

\usepackage{latexsym}
\usepackage{amsfonts, amsmath}
\usepackage{multirow}
\usepackage{color}
\usepackage{enumerate}
\usepackage{comment}
\usepackage{xspace}
\usepackage{booktabs}
\usepackage{enumitem}

\makeatletter
\def\url@leostyle{%
  \@ifundefined{selectfont}{\def\UrlFont{\sf}}{\def\UrlFont{\small\bf\ttfamily}}}
\makeatother
\def\pprw{8.5in}
\def\pprh{11in}

\PassOptionsToPackage{pdfpagelabels=false}{hyperref}
\usepackage[pdfencoding=auto]{hyperref}
\hypersetup{
pdftitle={SceneSuggest: Context-driven 3D Scene Design},
pdfauthor={Manolis Savva, Angel X. Chang, Maneesh Agrawala},
pdfkeywords={},
bookmarksnumbered,
pdfstartview={FitH},
colorlinks,
citecolor=black,
filecolor=black,
linkcolor=black,
urlcolor=black,
breaklinks=true,
}

\usepackage{cleveref}

\newcommand{\SceneSuggest}{\textsc{SceneSuggest}\xspace}
\newcommand{\none}{{\texttt{none}}\xspace}
\newcommand{\basic}{{\texttt{basic}}\xspace}
\newcommand{\full}{{\texttt{full}}\xspace}

\DeclareMathOperator*{\argmax}{arg\,max}

\newcommand{\lrtest}[4]{$\chi^2(#1,N=#2)=#3, p<#4$}

\teaser{
  \includegraphics[width=\linewidth]{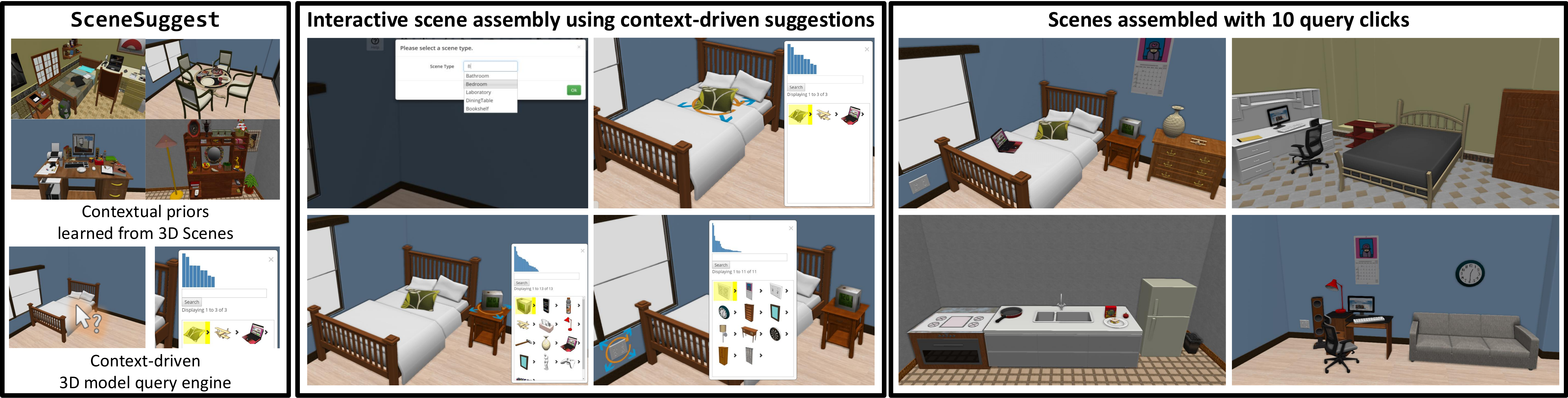}
  \caption{We present a context-driven 3D scene design system that provides intelligent contextual autocompletions.  Our system is based on a set of rich priors extracted from existing 3D scenes that enable a contextual 3D model suggestion engine (left).  Users specify a desired scene type and then indicate locations within the scene where models should be automatically suggested and placed (middle).  Scenes can be assembled with a small number of point-and-click contextual query operations (right). }
  \label{fig:teaser}
}

\begin{document}

\title{SceneSuggest: Context-driven 3D Scene Design}
\numberofauthors{1}
\author{
  \alignauthor Manolis Savva, Angel X. Chang, \textnormal{\large and} Maneesh Agrawala\\
  \affaddr{Computer Science Department, Stanford University}\\
  \email{$\{\textnormal{\large msavva, angelx, maneesh\}@cs.stanford.edu}$}\\
}
\toappear{}
\maketitle
\begin{abstract}
We present \SceneSuggest: an interactive 3D scene design system providing context-driven suggestions for 3D model retrieval and placement.  Using a point-and-click metaphor we specify regions in a scene in which to automatically place and orient relevant 3D models.  Candidate models are ranked using a set of static support, position, and orientation priors learned from 3D scenes.  We show that our suggestions enable rapid assembly of indoor scenes.  We perform a user study comparing suggestions to manual search and selection, as well as to suggestions with no automatic orientation.  We find that suggestions reduce total modeling time by 32\%, that orientation priors reduce time spent re-orienting objects by 27\%, and that context-driven suggestions reduce the number of text queries by 50\%.
\end{abstract}
\category{H.5.m.}{Information Interfaces and Presentation (e.g. HCI)}{Miscellaneous}

\section{Introduction}

Assembling 3D scenes end environments for interior design, product visualization, and games is hard.  Professionals train for extended periods of time to learn the byzantine frameworks that are prevalent in the 3D content creation industry.  The barrier to entry for incidental, occasional design of 3D content is large.  In contrast, assembly of 2D images and clipart is commonly performed by novices with no expertise.

One key factor separating the 2D and 3D realms is the degree to which composable, easy to retrieve content is available.  In 2D, clipart, photographs, illustrations and diagrams are ubiquitous.  In the latter, 3D models were until recently costly and hard to obtain.  The growth of public repositories such as the 3D Warehouse\footnote{\url{https://3dwarehouse.sketchup.com}} has changed this landscape significantly in the last decade.

With the improved availability of 3D models, the remaining bottleneck in scene design is in model retrieval and scene layout specification. The manipulations that are necessary to retrieve, place, and orient objects in order to compose a 3D scene require significant manual effort.  That is the case both with traditional input devices such as mouse and keyboard, and with more exotic higher DOF devices, which are less common and require additional familiarization. While recent work has looked at automatic generation of scenes, scene design is inherently an iterative process that is more well-matched to interactive systems.  

At the same time, there has been a recent explosion in interfaces for 3D scene assembly, predominantly in the context of interior design (e.g., SketchUp, Autodesk HomeStyler, and Planner5D).  A survey of twelve existing interior design systems revealed that all systems still use tried-and-true manipulation methods relying on repetitive translation, rotation and scaling operations in 3D.  Only two allow for retrieving models by text search while the rest present a hierarchical list for model selection.  Moreover, no existing systems attempt to use data-driven methods to suggest likely object placements, or to reduce the need for manual layout specification.

In this paper, we present \SceneSuggest: a 3D scene design interface that leverages existing data to power a contextual suggestion engine for rapid assembly of 3D scenes from a corpus of object models (see \Cref{fig:teaser} right).  Using data-driven learning methods we extract object support, position, and orientation priors from existing 3D content, and use them to make suggestions to users during 3D scene assembly (\Cref{fig:teaser} left).  We show that these priors can be used to predict object occurrence and positioning and reduce the number of operations required to assemble a scene.  We run a user study contrasting our contextual suggestions with keyword search and manual selection and find that overall modeling time is reduced by 32\% on average, and the number of text queries is reduced by 50\%.  The contextual suggestion interface reified in \SceneSuggest is the analogue of textual autocomplete for 3D scenes.

We make the following contributions in this paper:
\begin{itemize}[noitemsep]
  \item We present a system to generate contextual suggestions based on the position and orientation of 3D objects during interactive scene design
  \item We learn a set of priors targeted to contextual queries and show how they can be used to generate suggestions with 3D placement information
  \item We empirically evaluate the benefit of contextual suggestions in a formal user study and show that there is a significant reduction in scene modeling time and effort compared to selection from a fixed order list and keyword search
\end{itemize}

\section{Related Work}

Suggestive interfaces have been used successfully in systems text entry that learn from data and recency statistics~\cite{darragh1992reactive}, and more specifically in the context of translation systems~\cite{langlais2000transtype}.  User interfaces with autocomplete suggestions are now ubiquitous in many domains, including data visualization~\cite{koop2008viscomplete}.

For drawing and geometric modeling in 3D, reasoning purely with the geometry can assist drawing by guiding user interaction~\cite{igarashi2001suggestive,igarashi2007teddy}.  A similar line of work presents data-driven suggestion systems for 3D shape modeling~\cite{tsang2004suggestive,chaudhuri2010data}.

More recently, there has been work in context-based suggestive interfaces for 3D scene design.  Contextual queries for 3D scenes were introduced by Fisher and Hanrahan~\cite{fisher2010context} and Fisher et al.~\cite{fisher2011characterizing}.  However, this work only addressed model and scene retrieval, not scene design.  It was not integrated into a functional design system and was not evaluated in an interactive setting.

The ClutterPalette system~\cite{yu2014clutterpalette} learns static support priors from annotated RGB-D images using an approach similar to ours but they do not consider the continuous distributions over position and orientation that are our focus.  In addition, their system is tailored for detailing existing interiors.

Both our work and ClutterPalette draw upon two lines of prior work in scene design: procedural synthesis of 3D scenes, and interactive interfaces for 3D scene design.  The former requires the user to specify desirable properties of the output scene fully in the input, whereas the latter allows the user to interactively assemble 3D scenes.

\subsection{3D Scene Generation}

Procedural generation approaches have typically used manually specified rules and design principles.  Early scene layout generation systems were presented by Merrell et al.~\cite{merrell2011interactive} and Yu et al.~\cite{yu2011make}.  However, both assume that a pre-specified set of models is given and only optimize the layout.

More recently, Fisher et al.~\cite{fisher2012example} present a data-driven synthesis system which allows for tailoring to specific types of scenes, focusing on plausibility and variety of generated output.  In follow up work, Chang et al. synthesize 3D scenes from text~\cite{chang2014spatial}.  Both these systems are intended to generate scenes that are good starting points for further refinement.  However, their output is hard to control, and interactive manipulation which is our focus is outside their scope.

\subsection{Interactive 3D Scene Design}

Early work in interactive 3D scene design by Bukowski and Sequin~\cite{bukowski1995object} has demonstrated that associations of objects to surfaces through reasoning about physical support allow for more intuitive scene manipulation UIs.  Follow up work by Gosele and Stuerzlinger~\cite{gosele1999semantic} has shown that additional notions of object binding and offer areas representing typical static support patterns can lead to more efficient 3D scene design.  However, the semantics and priors on static support and object occurrence are assumed to be given as input.

Most of this prior work has been rule-based.  Data-driven methods have been largely unexplored with the exception of ClutterPalette.  The focus of this paper is to demonstrate how a richer set of support, position and orientation priors learned from 3D scene data can be leveraged for more efficient interactive scene design.


\begin{figure*}
  \includegraphics[width=\linewidth]{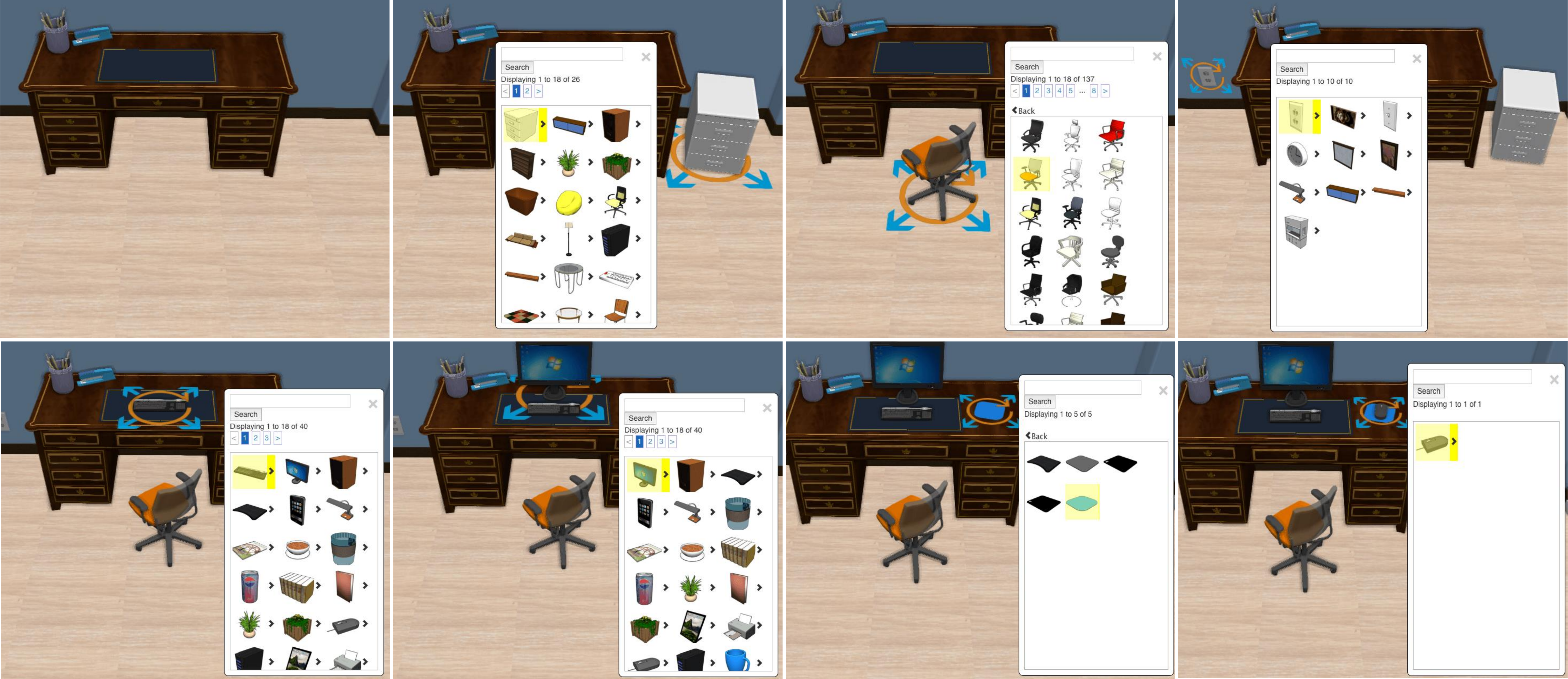}
  \caption{A sequence of contextual queries with automatically placed models and ranked alternatives.  Starting with the desk scene at the top left, a query by clicking on the floor to the right of the desk returns a filing cabinet (top mid left). Then a query in front of the desk returns a chair (top mid right). The subsequent queries retrieve: power socket on wall, keyboard on desk, monitor behind keyboard, mousepad to the right of the keyboard, and finally mouse on the mousepad.  All returned models are placed and oriented automatically by conditioning on the current context of the scene.}
  \label{fig:query-sequence}
\end{figure*}


\section{System Overview}

\subsection{Context-driven 3D Scene Design}

A user of the \SceneSuggest system starts with a partial or empty 3D scene (see \Cref{fig:query-sequence} top left).  They optionally specify a scene type to tailor the priors for more appropriate contextual suggestions.  During design, the user shift-clicks a point in the scene where they would like to add an object (e.g., floor next to desk to get filing cabinet in \Cref{fig:query-sequence}).

The system locates the surface point that was clicked and extracts a context query region anchored at that point.  The context query region includes information on the supporting parent surface, the support surface normal, and the current scene.  Using this information \SceneSuggest retrieves contextual priors and combines them in order to suggest a list of relevant 3D models.

Each suggestion consists of a category, placement (3D transform applied to the model), and score.  The result list is displayed to the user as a set of thumbnails in a floating panel next to the query point, and the top suggestion is automatically placed at the query point.  Other suggestions can be selected to replace the top one, or the list can be refined through text search.  Placed models are oriented automatically.

\subsection{Design Goals and Issues}

\SceneSuggest aims to reduce the manual manipulation effort in 3D scene assembly and make scene design easier for novices with no experience in using 3D content creation tools.  The goal of reducing low level manipulation effort implies a few key desiderata and corresponding design decisions:

\begin{itemize}[noitemsep]
\item{\emph{Requirement}: minimize number of user manipulation operations.  \emph{Design}: single click context-driven model placement, and support-based drag-drop manipulation.  Models are automatically oriented when inserted.}

\item{\emph{Requirement}: minimize need for manual annotation of objects with properties or constraints.  \emph{Design}: preprocess to learn priors from existing 3D scenes and leverage them during interactive assembly.}

\item{\emph{Requirement}: easy exploration of alternative objects, and manual manipulation when desired.  \emph{Design}: context-driven suggestion UI is coupled with traditional text-based keyword search, and widget for re-orienting objects.}
\end{itemize}

\subsection{Architecture}

We implement \SceneSuggest as a web-based client and server architecture.  This allows us to push computationally intensive queries onto the server while making the interactive frontend easily accessible from any web-connected device.

As a preprocess, we learn contextual priors from a corpus of 3D models and scenes.  We use the ShapeNetSem model dataset provided by Savva et al.~\cite{savva2015semgeo}.  In this dataset, models are categorized and aligned, and scaled automatically   \cite{savva2014sizes}.  Metadata such as model name (e.g., karlstad sofa), tags (e.g., modern, antique), and description (e.g., ``tall, leather office chair'') are associated with each model.  We index the models with the Apache Solr\footnote{\url{http://lucene.apache.org/solr/}} search engine for easy text queries.  Attachment surfaces (binding areas) and support surfaces (offer areas) for objects are learned from observations in scenes.  Our system only requires a category label and semantic up and front orientation for each model.  Most public repositories and annotated datasets provide this information, but this input requirement can be removed through automatic classification and 3D mesh alignment methods.

\begin{figure*}
  \vspace{-2em}
  \includegraphics[width=\linewidth]{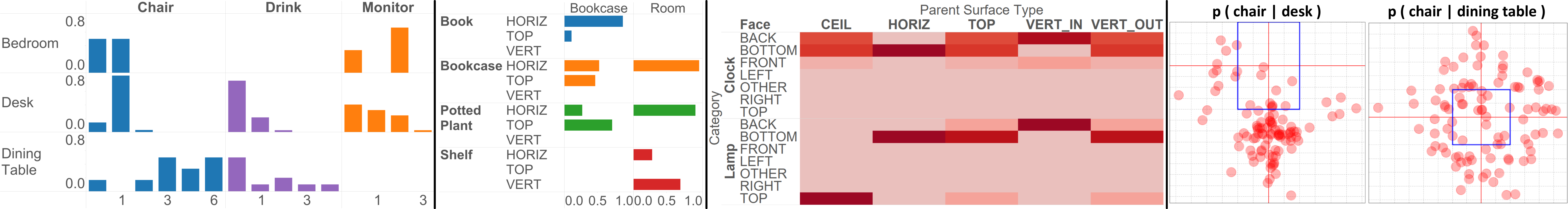}
  \caption{Examples of priors used by our system. Left: object occurrence count priors. Mid-left: support surface priors. Mid-right: attachment surface priors. Right: samples from position prior probability distributions.}
  \label{fig:priors}
  \vspace{-1em}
\end{figure*}

\begin{figure}
  \includegraphics[width=\linewidth]{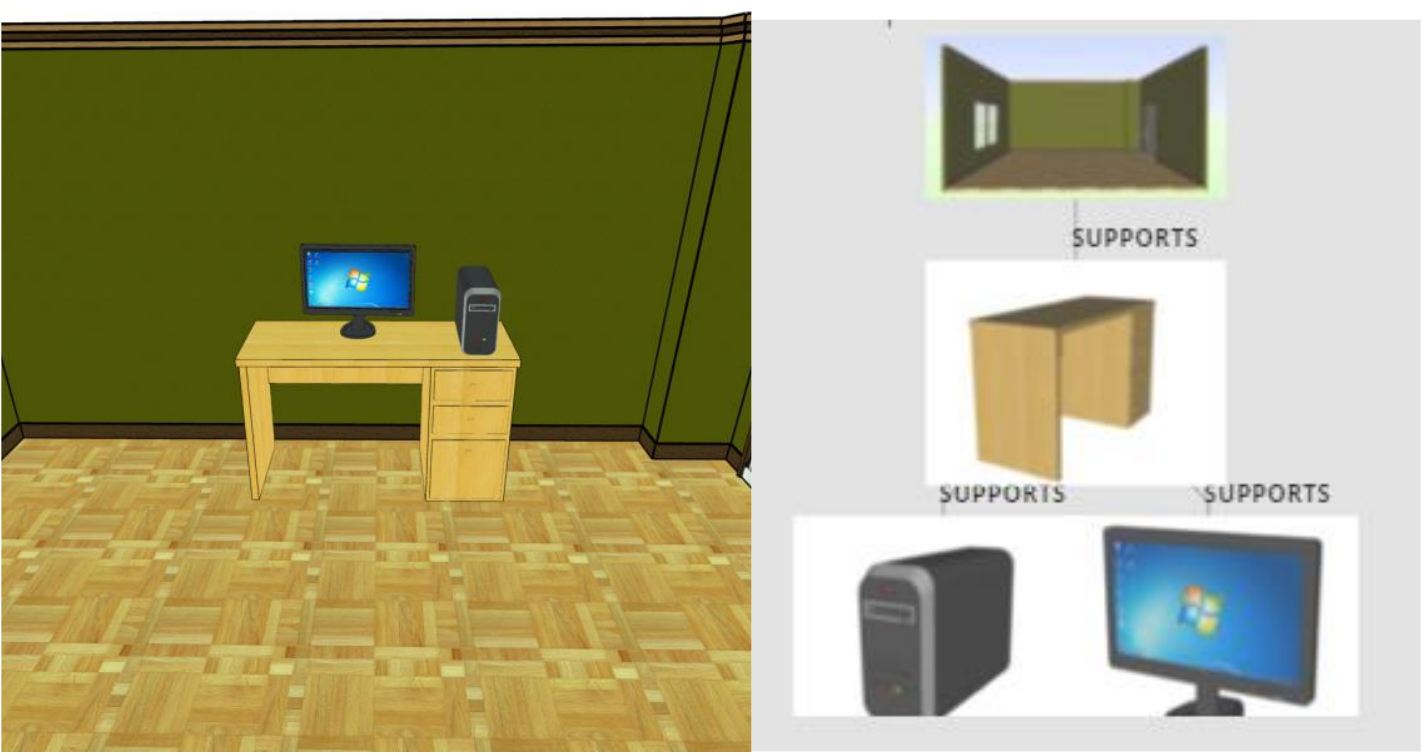}
  \caption{Example support hierarchy.  The ``Room'' supports the ``Desk'', which in turn supports the ``Monitor'' and the ``Computer''.}
  \label{fig:sceneHierarchy}
\end{figure}

We use the scene database from Fisher et al. \cite{fisher2012example}.  This dataset consists of approximately 130 simple indoor scenes of living rooms, bedrooms, bathrooms, and kitchens.  We use synthetic scenes rather than annotated RGB-D scenes since synthetic scenes contain rich contexual information in 3D space (e.g., relative orientations and distances).  Priors can also be obtained from annotated RGB-D scenes with a similar approach to the one we present, though this is beyond our scope.

We extract a set of contextual priors from the scenes: occurrence count probabilities of object categories given a scene type and support parent (e.g., number of monitor found in a living room), support and child attachment surface probabilities, and continuous probability distributions encoding relative distances and orientations between objects.  Variations of these kinds of priors have been introduced by prior work in scene analysis and synthesis \cite{fisher2012example,yu2014clutterpalette,chang2014spatial}.  Our focus is not to present a novel set of priors but rather to show how they can be leveraged in an interactive system for context-driven 3D scene assembly.  With the exception of \cite{yu2014clutterpalette} which has used conditional support priors for 3D scene detailing, the other priors have not been implemented or evaluated in interactive scene design systems.

\section{Learning Contextual Priors}

Our priors are learned from a corpus of 3D scenes composed of 3D object models.  We analyze the static support hierarchy and relative observations of models to obtain a set of contextual priors.  Here we discuss the scene representation and define the priors that we extract.

\subsection{Scene Representation}
A scene $s$ consists of a set of 3D model instances $\{o_1,\ldots,o_n\}$ where each model instance $o_i = (m_i, T_i)$ is a tuple containing a 3D model mesh $m_i$ from the model database and a transformation matrix $T_i$.  The model represents the physical appearance (geometry and texture) of the object, while the transformation matrix encodes the position, orientation, and scale of the object.  In addition, each scene has an associated scene type (e.g. ``bedroom'', ``living room'').

Static support relations between the objects are defined in a tree with an oriented edge $e_{ij}$ linking object $o_i$ to $o_j$ if $o_i$ is statically supported by $o_j$ (e.g., a bowl $o_i$ on a kitchen counter $o_j$).  The scene dataset we use includes a support tree hierarchy (see~\Cref{fig:sceneHierarchy}) for each scene.  Following Savva et al.~\cite{savva2015semgeo}, we identify the support and child attachment surfaces by considering all surfaces within a proximity threshold to the midpoint of each bounding box face around the supported object's bounding box plane.

\subsection{Contextual Priors}
We use the dataset provided by Savva et al.~\cite{savva2015semgeo}. Relative position and orientation priors are encoded following the approach of Chang et al.~\cite{chang2014spatial}.

We estimate the contextual priors using observations of categorized objects in the 3D scenes. To handle data sparsity we utilize the category taxonomy used by Savva et al. and back off to a parent category in the taxonomy for more informative priors if there are fewer than $k=5$ support observations of a given object's category.

\paragraph{Object Occurrence Counts}
Unlike prior work~\cite{fisher2012example,yu2014clutterpalette,chang2014spatial}, we model the probability of a object category being supported by a parent category in a given scene type directly.  In addition, we model the object count statistics. Data sparsity is addressed by using a backoff scheme using the category taxonomy.

We compute the probability of a child category $C$ given its support parent $p$ and the scene $s$ as context: $P(C | p,s)$.  We make the simplifying assumption that $P(C | p, s) = P(C | p_C, s_C, k)$ where $p_C$ is the support parent category, $s_C$ is the scene type, and $k$ is the number of existing support children on $p$ with category $C$ in scene $s$.  This allows us to model the cardinality of the expected number of instances for a given object category on a support parent (for instance, we would expect two speakers on a desk, and one keyboard).  Note that here we do not consider objects of other categories that may occur in the scene.

Then: {\small $P(C | p_C, s_C, k) = P(| C \text{~on~} p_{C} \text{~in~} s_{C}| > k | p_C, s_C)$.}
 
More concretely, we maintain a histogram for object category counts given the parent category $p_C$ and scene type $s_C$:

{\small $P(|C|=k|p_C, s_{C}) = \frac{\text{count}(| C \text{~on~} p_{C} \text{~in~} s_{C}| = k)}{\text{count}(C \text{~on~} p_{C} \text{~in~} s_{C})}$}

Which gives:

{\small $P(|C \text{~on~} p_{C} \text{~in~} s_{C}| > k | p_C, s_C) = \sum_{i > k} P(|C|=i|p_C, s_{C})$}

\Cref{fig:priors} left shows some examples.  Note that drinks do not occur in bedrooms in our scenes, and that the number of chairs is higher for dining rooms than for bedrooms.

\paragraph{Support and Attachment Surface}
The parent support surface priors are given by:

{\small $P_{\textit{surf}_\textit{sup}}(t | C) = \frac{\text{count}(C \text{~on surface with~} t)}{\text{count}(C)}$}

The parent supporting surface is featurized using the surface normal (up, down, horizontally) and whether the surface is interior (facing in) or exterior (facing out).  For instance, a room has a floor which is an upwards interior supporting surface. 

The child attachment surface priors are given by:

{\small $P_{\textit{surf}_\textit{att}}(t | C) = \frac{\text{count}(C \text{~attached at surface~} t)}{\text{count}(C)}$}

Object attachment surfaces are featurized using the bounding box side: one of top, bottom, front, back, left, or right.  For instance, posters are attached on their back side to walls, rugs are attached on their bottom side to floors.

If there are no observations available we use the model geometry to determine the support and attachment surface. For support surfaces we pick only upward facing surfaces, while for attachment we assume 3D (blocky) objects are attached on the bottom (e.g. paper boxes), flat objects are attached on their back or bottom (e.g. posters), and thin objects are attached on their side (e.g., pens).

\Cref{fig:priors} middle show examples of support surface and attachment priors.  Note how books are mostly found on horizontal surfaces but not the top of bookcases, whereas potted plants tend to go on top of bookcases.  We also see that some bookcases are stacked.  For attachment surface priors, clocks and lamps exhibit different face probabilities for different support surfaces.  If these categories are broken down into subcategories (e.g., wall lamp, desk lamp, etc.) then the attachment priors tend to be favor one face.

\paragraph{Relative Position and Orientation}
We model the relative positions and orientations of objects based on their object categories and current scene type: i.e., the relative position of an object of category $C_\textit{obj}$ is with respect to another object of category $C_\textit{ref}$ and for a scene type $s_C$.  We condition on the relationship $R$ between the two objects, whether they are siblings ($R=\textit{Sibling}$) or a child-parent pair ($R=\textit{ChildParent}$).  

In addition, we also condition on the support surface $t$ of $C_\textit{obj}$.  We project the centroids of the two objects onto the support surface, and use the offset in that plane $\delta = (x,y)$ as the relative position.  For reference objects that do not have a semantic front (e.g. circular objects like round tables), we represent the delta as the radius from the center.

To summarize, we define the relative position prior as: $P_\textit{relpos}( \delta,\theta | C_\textit{obj}, C_\textit{ref}, s_C, R, t )$.  For simplicity, we assume the relative position and orientation are independent.  We model the relative position as a mixture of multivariate gaussians and estimate the parameters using kernel density estimation (see \Cref{fig:priors}).  The figure shows centroid position samples drawn from one category (red points) being conditioned on the presence of another category (blue outline).  For encoding relative orientations, we use a wrapped histogram binned into 36 bins of 10 degrees each.

\begin{figure}
  \includegraphics[width=\linewidth]{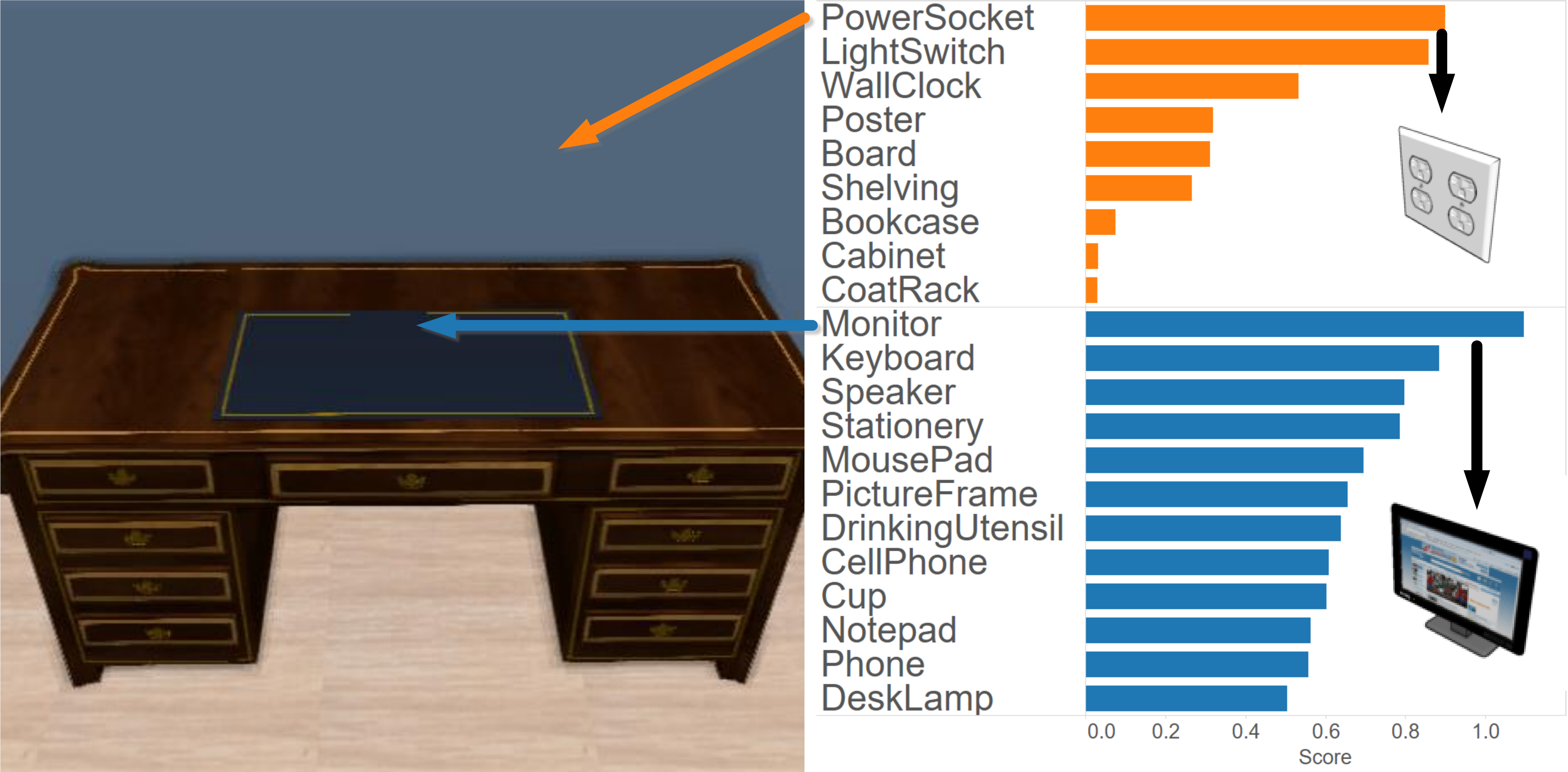}
  \caption{Conditional probabilities of supported object categories for two points in a scene with a desk: objects attached on the wall above the desk (orange), and objects on the top of the desk (blue).  The distributions have been truncated for presentation---there is a long tail capturing a variety of categories that can be supported by each region.}
  \label{fig:desk-scores}
\end{figure}

\begin{figure*}
  \centering
  \vspace{-2em}
  \includegraphics[width=\linewidth]{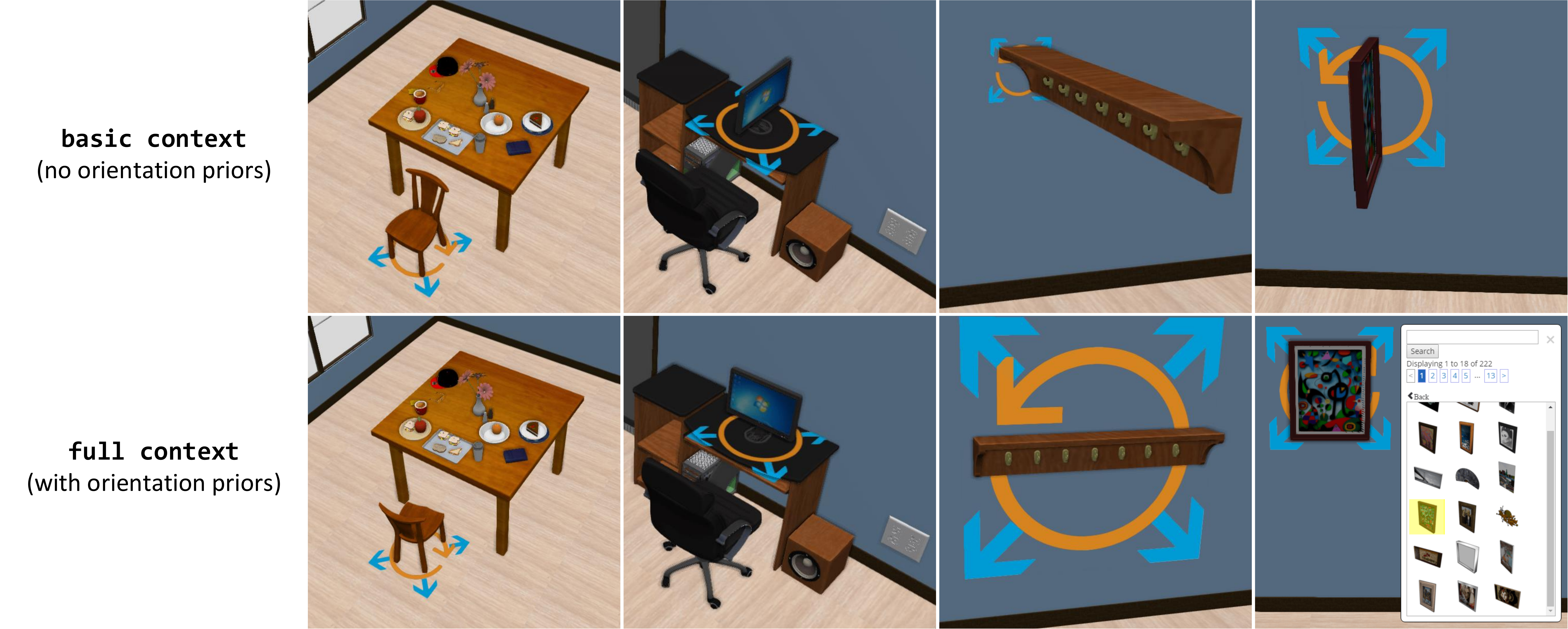}
  \caption{Comparison of contextual suggestions with \basic context priors and with \full context priors.  From left to right: chair in front of table, monitor on desk, coat rack on wall, poster on wall.  Note how the \full model determines reasonable relative orientations for the suggested models.  The sides of the objects which are in contact with the attachment surface, and their upright and front orientation are predicted by our system.}
  \label{fig:basic-vs-full}
\end{figure*}

\begin{figure}
  \includegraphics[width=\linewidth]{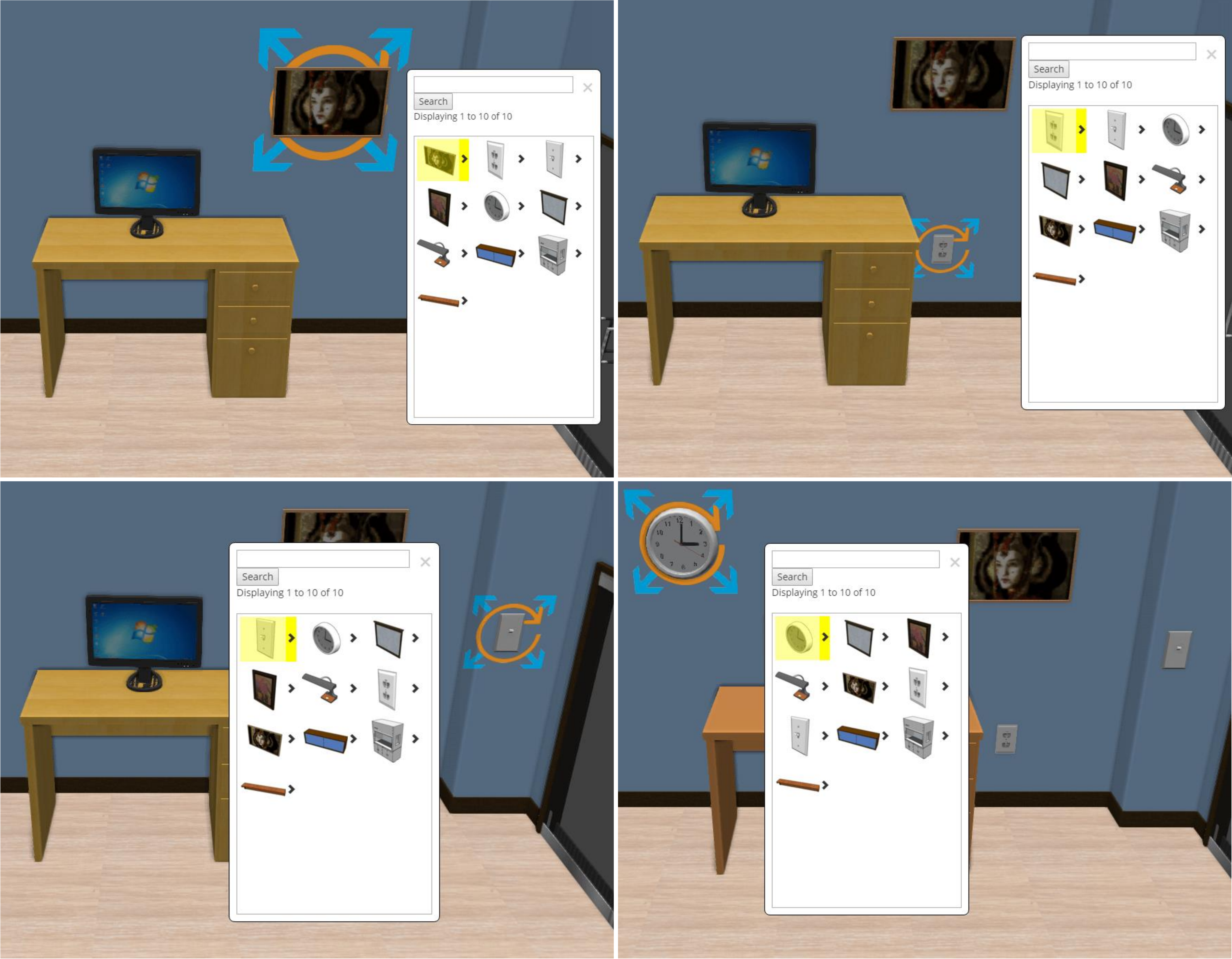}
  \caption{Contextual queries at different positions on a wall.  From top left: poster at top, socket at bottom next to desk, switch at arm height by the door, and clock high above the desk.  The probabilities of different categories vary with height and relative position from other objects.}
  \label{fig:wall-positions}
  \vspace{-2em}
\end{figure}

\section{Generating Contextual Suggestions}

After the learning preprocess is completed for a given input 3D scene corpus, the learned priors are encoded in a web service that the server component of \SceneSuggest can retrieve on demand.  The counterpart client component consists of an interactive WebGL-based 3D design UI which makes calls to the server whenever contextual queries are performed.  

When the user starts a query by clicking in the scene, they are implicitly specifying a context query region.  We define this context region query to be $R = (s, p_C, p_N, t, pos)$ where $s$ is the current scene, $p_C$ is the supporting parent object category, $p_N$ is the normal at the point on the supporting parent object's surface, $t$ is the supporting surface type, and $pos$ is the 3D position of the anchor point on the surface.  Given a user click we determine these values by raytracing into the 3D scene.  This context region is streamed to the server where a corresponding scene proxy is recreated and used for computing relevant priors.

The context query returns a ranked list of model placement suggestions.  Each suggestion $S = (C,M,w)$ consists of an object category $C$ and a placement $M$, and a score $w$.  The placement is represented as $M = (T,F)$ where $T$ is a $4\times4$ transformation matrix, and $F$ indicates the child attachment face (side of child object's bounding box in contact with parent).  The matrix $T$ specifies the position, orientation, and scale of the object.  Since position is provided as input in the context query region, and size is assumed to be fixed, the client UI uses only the orientation to automatically orient placed objects.

We compute the overall score $w$ for ranking suggestions as a linear combination $w = \lambda_1 P(C | p_C,s) P_{\textit{surf}_\textit{sup}} (t | C) + \lambda_2 w_\textit{pos}$. We used $\lambda_1=1, \lambda_2=0.25$ for all presented results.  This term combines the probability that the category $C$ is supported by the parent category $p_C$ with the probability that the selected support surface is appropriate for the given category.  In addition, we take into account the position of the object using $w_\textit{pos} = \sum_{o_j \in F(o_i) }P_\textit{relpos}(\cdot)$ where $F(o_i)$ are the sibling objects and parent object of $o_i$.  The orientation is obtained by first determining the child attachment face $F$ for the given support surface type, and then computing the rotation that orients $F$ toward the supporting surface.  $F = \argmax_{f} P_{\textit{surf}_\textit{att}}(f | t)$ is the most likely child attachment face for the surface type $t$.  Finally, we pick a rotation angle $\alpha$ around the support surface normal that gives highest $w_\textit{pos}$.

\Cref{fig:desk-scores} illustrates the ranked suggestion lists returned by the server in response to two different contextual queries (query point on the wall behind the desk vs. query point on the desk).

Once the ranked list is returned to the client UI, the list is displayed to the user as a floating panel with clickable object thumbnails.  The first object is automatically oriented and placed at the query point.  In the suggestion list, each category of objects is represented by the thumbnail of a representative model.  We chose to group the models by category in order to present an initial list that is clean but varied. From the initial list, the user can drill down to expand a category (by clicking a right pointer icon), thus exploring model instances for a given category. Currently, our method gives the same score and orientation for all models in the same category group.  An interesting avenue for future work would be to take into account instance size and style to assign different scores for different instances of an object category.



\section{Results}

Figure~\ref{fig:teaser} illustrates how \SceneSuggest can be used to rapidly assemble several different types of scenes.  One of the key features of our system, is the automatic orientation of objects.  Without pre-annotated semantic information about how objects are typically oriented with respect to each other, the user needs to manually re-orient placed objects (\Cref{fig:basic-vs-full} top).  Using the learned priors, \SceneSuggest automatically orients the suggested objects (\Cref{fig:basic-vs-full} bottom).

While it is possible to manually specify for each model, the offer and binding areas as in Gosele and Stuerzlinger~\cite{gosele1999semantic} (or specify general rules predicated on categories), we are able to learn these orientation priors directly from data. This reduces the need for manual annotation and allows us to easily incorporate new model instances in our system.

Furthermore, our data-driven approach can learn more subtle and harder to annotate spatial priors between objects (e.g., the most likely object in front of a desk's leg space is a chair, and it should be oriented toward the desk's leg space).  This form of subtle spatial prior is demonstrated in \Cref{fig:wall-positions}.  Given the same support object and same support surface (wall), the object categories suggested are ranked differently depending on the selected position.  Posters and clocks are typically found high on the wall, a light switch is at arm height, and power sockets are found near the bottom of the wall.

\begin{figure*}
    \includegraphics[width=\linewidth]{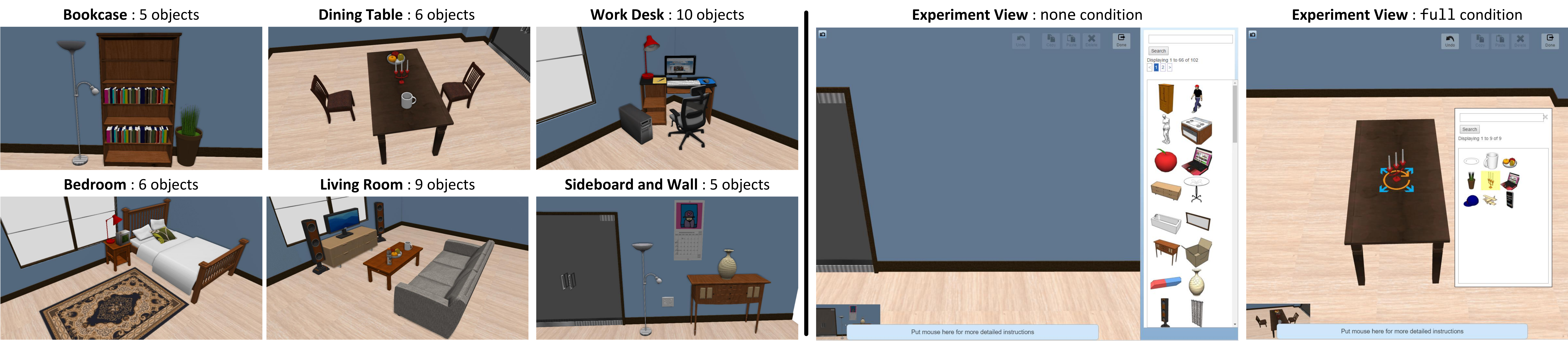}
    \caption{Left: the six target scenes in our user study.  The scenes ranged in object complexity (5-10 objects) and need for complex re-orientations (e.g., putting posters and sockets on the wall). Right: view of the interface seen by participants during a \none condition scene assembly and during a \full condition scene assembly.  The user can enlarge the target scene at the bottom left or view detailed instructions at the bottom by hovering.}
    \label{fig:experiment}
\end{figure*}

\section{User Study}

We carried out a user study to evaluate the context-driven suggestions in the \SceneSuggest interactive scene assembly system.  The study task was to recreate a shown scene by selecting, placing and orienting 3D models of objects.

We created a set of six target 3D scenes representing common indoor object arrangements (see \Cref{fig:experiment}) using a traditional scene design interface.  We then selected a representative model from each of the top 100 categories in the model corpus of \cite{savva2015semgeo} to create a corpus of 102 models (the models in the experiment target scenes were required to be representatives for their category, leading to more than 100 models due to two chairs and tables).  During the scene assembly task users selected from these models to recreate the target scenes.

\subsection{Hypotheses}
Based on the design goals of our system and informal testing, we predicted that:

\textbf{H1.} Assembling the target scenes should be faster with context-driven suggestions than with keyword search and selection from a list.  We expect that relevant suggestions for objects will reduce user effort in finding and inserting objects into the scene.

\textbf{H2.} Scene assembly will be faster with full priors (including orientation information in the context) than with basic priors (excluding orientations).  The automatic orientation of objects within the context indicated by the user should reduce the time spent in re-orienting objects manually.

\textbf{H3.} Total time spent re-positioning objects will be lower with context-driven suggestions compared to suggestions with no automatic object orientation.  Direct placement of the objects in their desired context using contextual suggestions should reduce the need for translation operations to change the placement of the objects.

\subsection{Methods}
\subsubsection{Participants}
We recruited 20 participants through Amazon Mechanical Turk (12 male, 8 female, range of 22-57 years old, average age of 33 years) .  Participants were required to be fluent speakers of English and reside in the United States.  Five of the participants reported some experience using 3D design UIs such as SketchUp before the study, while the rest did not have any prior experience.  Participants were compensated with 5 USD for performing the study.

\subsubsection{Design}
The experiment was contrasting three interface conditions: \none, \basic, and \full. The \none condition was a default interface with no contextual suggestions---instead, a search panel displaying all the available models in fixed ordering was provided.  Users could also search for models by keyword using a textbox at the top of the panel.  The \basic condition provided contextual suggestions for models by ``shift-clicking'' at any point within the 3D scene.  The suggestions were again shown in a search panel with a search textbox at the top so the user could override the suggested list with a manual keyword search.  However, the models were ordered by descending probability conditioned on the basic categorical and positional priors.  The top ranked model was inserted by default at the chosen anchor point and the user could click on any other model to place at the same position instead.  The \full condition retains the same interface but the suggestions are additionally automatically oriented at the anchor point using the learned orientation priors.

The three conditions were contrasted in a within-subject design for three target scenes (each condition-scene pair occurring once).  There were 3 conditions $\times$ 3 scenes = 9 trials per participant.  The presentation order of the trials was counterbalanced to control for learning using a balanced latin square design.  Participants were instructed in each trial to either select from the fixed ordering search panel, or to initiate a contextual suggestion by clicking in the scene.

There were 20 participants $\times$ 9 trials for a total of 180 assembled scenes.  The study was conducted in two periods separated by a day with the first period using the first three target scenes, and the second period using the other three.

\begin{table*}
    \vspace{-2em}
    \centering
    \begin{tabular}{lccc|cc}
        \toprule
        condition & total time [s]   & rotation time [s]  & translation time [s]  & model queries [count] & model query MRR \\
        \midrule
        \none     & 204 (177 - 241)  & 23.6 (19.9 - 28.7) & \emph{40.5 (32.0 - 51.1)}    & 5.2 (3.8 - 7.0) & 0.353 (0.318 - 0.388) \\
        \basic    & 170 (130 - 267)  & 22.4 (19.5 - 26.5) & \emph{38.5 (31.0 - 50.1)}    & 3.3 (2.3 - 4.9) & 0.785 (0.765 - 0.805) \\
        \full     & 139 (118 - 168)  & 17.2 (14.3 - 21.1) & \emph{39.8 (30.1 - 62.5)}    & 2.5 (1.6 - 3.9) & 0.769 (0.747 - 0.791) \\
        \bottomrule
    \end{tabular}
    \caption{Mean timings in seconds for scene assembly operations (and 95\% confidence intervals computed by bootstrapping with 1000 samples).  The \basic and \full contextual suggestion conditions reduce total modeling time significantly.  The \full model reduces the average time spent rotating models.  There is no significant effect of the condition on average translation times.  The average number of manual model search queries are significantly reduced by \basic and \full, and the mean reciprocal rank of the chosen models is significantly higher for both compared to the \none condition.}
    \label{tab:timings}
\end{table*}

\begin{figure}
  \vspace{-.5em}
  \includegraphics[width=\linewidth]{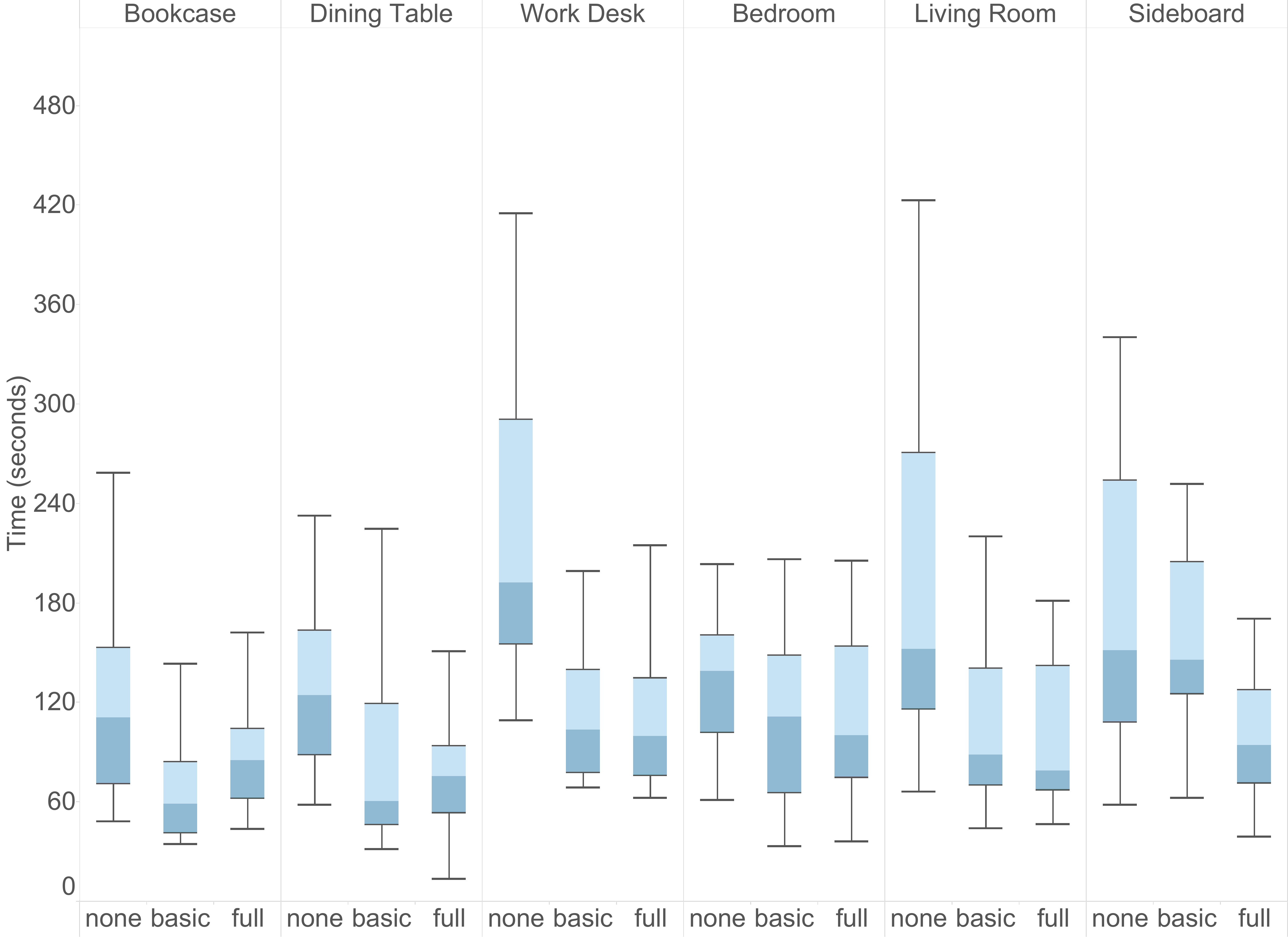}
  \caption{Box-and-whisker plots showing the medians and interquartile ranges for the total scene assembly time by condition for each of the six target scenes in our user study.  Overall, the \full and \basic conditions reduce modeling time significantly compared to the \none condition.}
  \label{fig:scene-times}
  \vspace{-.5em}
\end{figure}

\subsubsection{Procedure}
Before the start of the study participants were given a short description of the experiment and told that its goal was to contrast new UIs for 3D scene design.  The participants were told that they would see nine target images displaying a scene that they should assemble using available 3D objects.  We also informed the participants that sometimes they would be using contextual search by clicking in the scene, and that sometimes they would use a normal search panel to select and manually place objects.

The instructions briefly described other basic operations in the interface including translation of objects on support surfaces by click-dragging the object itself, and rotation of the object around its vertical axis by dragging a ring manipulator around the object (or through keyboard shortcuts).  The participant was asked to match the appearance of the target scene as efficiently as possible, but they were not given specific time goals.  At the end of the instructions, demographic data was collected in a short survey.

\begin{table}
\centering
\begin{tabular}{lcccc}
\toprule
condition / rank   & 1      & 2      & 3      & 4+ \\
\midrule
\none  & 29.6\% & 14.0\% & 0.35\% & 56.0\% \\
\basic & 68.7\% & 11.1\% & 4.28\% & 15.9\% \\
\full  & 67.5\% & 9.84\% & 4.87\% & 17.8\% \\
\bottomrule
\end{tabular}
\caption{Distribution of ranks of selected models for each condition.  Higher ranked suggestions are selected much more frequently in the \basic and \full conditions compared to the \none condition.  This is despite the fact that in the \none condition users predominantly specified what models they desired through text search.}
\label{tab:ranks}
\vspace{-1em}
\end{table}

Once the trials started, the participant would see a target image of the desired scene in the bottom left of their screen (hovering over the image allowed for zooming in to reveal detail).  In the main scene view, the participant proceeded to insert, move, and rotate objects until the target image was matched to their satisfaction (see \Cref{fig:experiment}).  Reminder instructions about the current interface were available on demand at the bottom of the screen by hovering on a ``more instructions'' message box.  When satisfied with the current scene, the participant would click a ``done'' button to move to the next scene.  All user interactions were transparently logged and timestamped during each trial, and recorded for later analysis.

After all 9 scene-condition pairs presented, an exit survey asked for subjective evaluations of enjoyment and competence for the default ordering panel, and the contextual suggestion interface on a 5 point Likert scale (1 being ``very low'' and 5 being ``very high'').  Finally, participants could indicate a preference for either the fixed ordering search panel, or the contextual suggestion interface.  They were not aware of the contrast between \basic and \full conditions for the context placements.

\subsection{Study Results}

The results of the study confirmed H1 (total modeling time reduction for \full and \basic) and H2 (total re-orientation time reduction in \full vs \basic) but not H3 (total re-positioning time reduction for \full and \basic).

We analyzed the overall time taken by users to match each target scene, as well as the time spent performing object translation (move) and re-orientation (rotate) operations (see \Cref{tab:timings} left for a summary).  We found that users spent an average of 3.4 minutes to assemble a scene under the \none condition.  Using the \SceneSuggest contextual suggestions, the mean modeling time was reduced to 2.8 minutes for the \basic condition and to 2.3 minutes for the \full condition (a reduction of mean scene assembly time by $32\%$).  Total scene assembly time exhibited high variance in all conditions and scenes but overall there is a significant reduction in modeling time from \none to \basic and from \basic to \full (see \Cref{fig:scene-times} for a breakdown by target scene).

We use a mixed effects model to account for the per-participant and per-scene variance on the total time (which is not accounted for by standard ANOVA).  We treat the participant and scene as random effects with varying intercept, and the condition factor as the fixed effect\footnote{We used the \texttt{lme4} R package and optimized fit with maximum log-likelihood. Results reported using the likelihood-ratio (LR) test.}.  We found that there was a significant effect of the condition factor on the total time: \lrtest{2}{180}{2807.8}{0.05}.  We also found a significant effect of condition on the total object rotation time: \lrtest{2}{180}{1429.1}{0.05}.  The reduction in re-orientation effort between \basic and \full confirms H2 and that automatically suggested orientations are useful to users.  We did not find a significant effect of the condition on the time spent translating objects.  This indicates that users translated models through drag and drop operations even when using contextual queries.  Unfortunately, we did not measure the total distance that users moved objects during scene assembly.

The total scene assembly times were significantly higher than those we observed in informal tests with collaborators.  We hypothesize that part of this could be due to not controlling for the user input device configuration (we did not require that participants use a mouse---some participants indicated that they used laptops with trackpads instead in the optional comments).  In addition, several participants commented that it was frustrating and difficult to rotate objects into appropriate orientations.  This is partly due to our rotation widget design which only allowed rotation around a single axis at a time (objects had to be re-positioned on different support surfaces to switch axes of rotation).

We also tracked the number of text search queries that users issued during each session.  Using \SceneSuggest, we were able to reduce the need for querying by $50\%$ (from an average of $5.2$ queries per session to $2.5$).  Additionally, higher ranked suggestions were selected in the \basic and \full conditions even when manual text searching was included (see \Cref{tab:ranks}).  We used the mean reciprocal rank (MRR --- a common measure of retrieval performance) for each selected object to evaluate how good the quality of the ranked suggestion lists.  As expected, the \texttt{basic} and \texttt{full} conditions have significantly higher MRR than the \texttt{none} condition.

Participants rated the contextual query conditions as more enjoyable (average enjoyment rating of 4.17 vs 3.23 for \none), and indicated that they felt more competent when using contextual queries (average competency rating of 4.38 vs 3.88).  Though freeform comments in the exit survey were optional, nine out of our twenty participants commented that they found the contextual query interface to be intuitive and preferable to searching through a fixed list.


\begin{figure}
  \includegraphics[width=\linewidth]{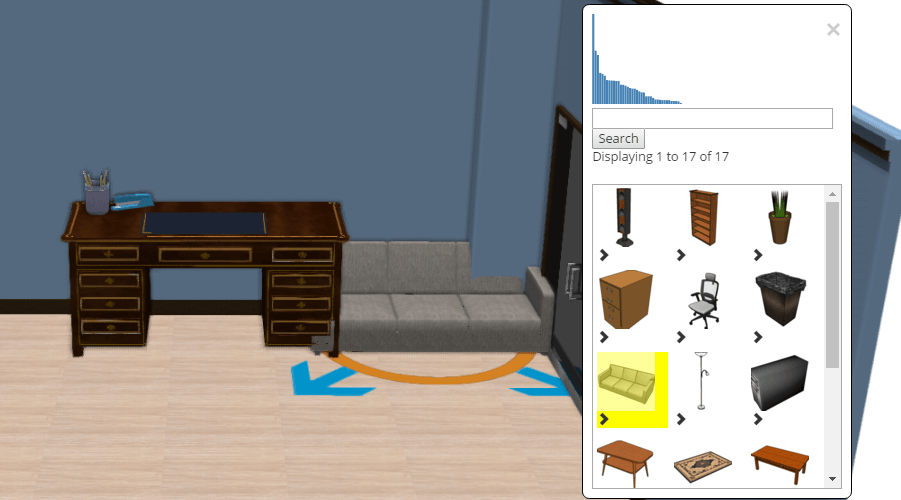}
  \caption{Our contextual queries do not take into account object sizes and do not reason about collisions.  This can result in failure cases such as this example of a large couch in the corner of the room.}
  \label{fig:size-limitation}
\end{figure}

\section{Discussion}

The results of our user study demonstrated that 3D scene design with the \SceneSuggest contextual query interface can lead to more efficient and more enjoyable assembly of scenes.  The direction of leveraging scene context information to retrieve relevant models and assist assembly is powerful and only partially explored in this paper.

\subsection{Limitations}

Currently our suggestion engine does not take into account the size of the suggested object, nor does it account for collisions.  \Cref{fig:size-limitation} shows a case where a couch selected by the user is too large to fit at the query point.  Some simple reasoning using the size of the object model along with collision checking can allow the system to suggest a different orientation, or to indicate that the placement is not feasible.

A related limitation of our system is that we do not allow for deviation from the anchor point that the user has specified.  Since the clicked location is likely to be a rough suggestion in many cases, adjusting the placement point to increase likelihood under the contextual priors may lead to better placements.

Another limitation of our system is that we do not model style compatibility for the objects within a scene, or address other long range, more than pairwise relations explicitly.

\subsection{Future Work}

A direct extension of the \SceneSuggest system could incorporate priors on style and compatibility to better model the aesthetics of a scene region.  For instance, beyond suggesting that chairs should go around a dining table, suggesting appropriate chairs that match the existing decor and color theme are likely to lead to better model suggestions.  Incorporating long distance dependencies such as symmetries in the placement of speakers next to TVs can enable multiple contextual placements to be jointly suggested (e.g., suggesting a set of chairs around the table).

Suggestion ranking could be significantly improved by incorporating history and personalization statistics that are commonly used by recommender systems.  Allowing for faceted ranking  of models using different scoring dimensions (e.g., size, popularity, style) could similarly be very helpful.


In our system we specified queries through a point and click metaphor.  Allowing users to mark a surface region in 2D, or to specify a 3D bounding box is an interesting direction.  Specification of bounding boxes or surface regions in 3D requires more complex interactions than just clicking so we did not explore this direction.  However, integrating the size and orientation information that such widgets provide can allow for more refined contextual queries.

Finally, tailoring the priors used by our system for suggestions currently requires that the user manually specify the type of scene they are interested in designing.  Automatically inferring scene type as objects are placed into the scene, or even inferring more refined types for regions within a scene is another interesting direction for future work.

\section{Conclusion}

We presented \SceneSuggest: a contextually-driven interactive 3D scene design system.  We showed how priors on the stucture of 3D scenes could be extracted from data and leveraged to offer real-time suggestions for model placements.

We empirically evaluated the contextual suggestions of our system against a simple baseline condition using a fixed order list and traditional keyword search for model retrieval.  The results of our user study indicate that total modeling time is reduced by the \SceneSuggest contextual suggestions, and that automatic object orientations further reduce modeling time.

Contextual queries during 3D scene assembly are a powerful tool for interactive design of scenes.  In this paper, we have barely scratched the surface of what is possible.

Along with the rising ubiquity of RGB-D sensing, VR and AR technologies, 3D scene data will continue to grow.  Data-driven methods for interactive 3D scene design will most likely prove to be a rich area for future research.

\balance
\bibliographystyle{acm-sigchi}
\bibliography{smartscenes}

\end{document}